\documentclass{article}

\usepackage{PRIMEarxiv}

\usepackage[utf8]{inputenc} 
\usepackage[T1]{fontenc}    
\usepackage{hyperref}       
\usepackage{url}            
\usepackage{booktabs}       
\usepackage{amsfonts}       
\usepackage{nicefrac}       
\usepackage{microtype}      
\usepackage{lipsum}
\usepackage{fancyhdr}       
\usepackage{graphicx}       
\graphicspath{{media/}}     
\usepackage{natbib}
\usepackage{amsmath} 
\usepackage{lineno}
\usepackage{subcaption, float}
\usepackage{placeins}
\usepackage{multirow}
\usepackage{xcolor}

\title{A dataset of one-dimensional idealized probabilistic fields
}

\author{
  Gregor Skok\\
  University of Ljubljana, Faculty of Mathematics and Physics\\
  Jadranska Cesta 19, 1000 Ljubljana,  Slovenia \\
  \texttt{Gregor.Skok@fmf.uni-lj.si} 
  \AND
  Romain Pic \\
  ETH Zurich, Seminar for Statistics, Department of Mathematics\\
  Rämistrasse 101, 8092 Zurich, Switzerland
   \\
   \texttt{romain.pic@math.ethz.ch} \\
}

\begin{document}
\maketitle


\begin{abstract}
Verification of probabilistic weather forecasts remains a crucial aspect of numerical weather prediction, as new AI-based models become more widely used alongside the more traditional physics-based ensemble forecasting systems that continue to be developed and improved. We present a first-of-its-kind idealized probabilistic dataset composed of one-dimensional cases aimed at analyzing the behavior and properties of verification methods for probabilistic forecasts and comparing their behavior. It covers a wide range of probabilistic cases, such as constant, localized events, gradients, fronts, noisy, bimodal, and limiting cases. Moreover, the code associated with the dataset provides great flexibility for customizing the experiments it covers. The dataset represents the first building block of the more extensive comparison dataset of the Bridging The Gap project, which aims to facilitate the development and comparison of spatial verification methods for probabilistic forecasts.

\end{abstract}


\section{Introduction}


Verification of weather forecasts remains a crucial aspect of numerical weather prediction, as new AI-based models become more widely used alongside the more traditional physics-based models that continue to be developed and improved. Numerical weather forecasts take the form of high-dimensional spatial fields, which are often characterized by strong correlations between nearby locations (i.e., spatial autocorrelation). This makes spatial verification, which accounts for the presence of spatial autocorrelation, crucial. Spatial forecast verification as a research field gained interest when non-spatial methods based on grid-point-by-grid-point verification were unable to highlight improvements caused by an increase in resolution of the models \citep{Gilleland2009}. The \textit{double-penalty effect} has crystallized the limitations of non-spatial verification methods \citep{Ebert2008}. The double-penalty effect refers to the fact that if a forecast presents a spatial displacement with respect to observations with no overlap between the two, the error would be penalized twice: once where the event was observed and again where the forecast predicted it. An undesirable consequence of this is that not forecasting the event is preferred to forecasting it slightly displaced. Moreover, non-spatial methods can also struggle to distinguish between near hits and significant spatial displacements \citep{Brown2011, Skok2022, Skok2025}.

Numerous spatial verification methods have been developed over the years to account for the spatial characteristics in the forecast and observation fields. However, these methods were initially proposed independently, and there was no central framework for their development or comparison. Then, the Spatial Forecast Verification Methods Intercomparison Project \citep[ICP;][]{Gilleland2009} was initiated to better understand the rapidly increasing literature on spatial verification methods. ICP was extended into a second phase, called Mesoscale Verification Intercomparison over Complex Terrain \citep[MesoVICT;][]{Dorninger2018}, with a focus on the effects of complex terrain, observation uncertainty, and spatial verification of probabilistic forecasts. Both projects heavily relied on comparison datasets to compare spatial verification methods and investigate their limitations and specificities.

Comparison datasets have played a key role in these meta-verification projects. The comparison datasets of ICP and MesoVICT included synthetic data, perturbed forecasts from real-world applications, and real-world datasets \citep{Ahijevych2009, Dorninger2018}. These datasets enabled a systematic and rigorous comparison of the existing spatial verification methods. While real-world and perturbed datasets are certainly important, the synthetic datasets provide a controlled environment for understanding and comparing the behavior of the methods. They allow us to isolate sensitivities to misspecifications, as we know which misspecification is present and how intense it is (e.g., if only displacement is introduced, and by how much and in which direction). This controlled environment can also enable us to compare and study verification methods in terms of operationalization, such as computational speed or resilience to pathological cases. Furthermore, the simpler setting it provides allows checking whether the results align with subjective evaluation.

ICP's synthetic dataset consisted of idealized elliptical precipitation patterns with one observation and five forecast fields \citep{Ahijevych2009}. MesoVICT expanded its synthetic dataset beyond elliptical cases to include pathological, circular, noisy, and scattered cases \citep{Gilleland2017, Gilleland2020}. Outside of these projects, \cite{Skok2022} additionally proposed a synthetic dataset based on geometrical cases that also include squares, Gaussian events, non-binary elliptical cases, and gradients.

However, all of the idealized synthetic comparisons described above were formulated as deterministic cases, and no analogous probabilistic dataset is currently available for evaluating verification methods tailored to probabilistic forecasts. This constitutes a clear knowledge gap, particularly as probabilistic forecasts provide increasingly valuable information and are becoming more widely available with the rapid development of AI-based forecasting systems \citep[see, e.g., ][]{Price2025, Lang2026}, along with the improvements of existing physics-based ensemble systems, that move to ever-higher resolutions. Despite MesoVICT's efforts to facilitate the development of spatial verification methods for probabilistic forecasts, these methods are underdeveloped \citep[see, e.g.,][]{Pic2025}. During the 2020 International Verification Methods Workshop, ``a lack of representation of novel spatial verification methods for ensemble prediction systems was noted'' \citep{Casati2022}. 

To address this gap, we propose a novel idealized  synthetic probabilistic dataset that can be used to evaluate the behavior of spatial verification methods for probabilistic forecasts. As the probabilistic aspect of the spatial fields increases their complexity, we consider one-dimensional spatial fields to facilitate visual representation and follow the simple nature of the deterministic geometric cases mentioned above.

The remainder of this article is organized as follows. Section~\ref{sec:description} describes the dataset and the different fields composing it. Section~\ref{sec:examples} showcases how the proposed dataset can be leveraged to compare spatial verification methods.

\section{Dataset description}\label{sec:description}

The dataset consists of 31 idealized one-dimensional situations. Each situation is represented as a two-dimensional field with dimensions 101 $\times$ 101 grid points, with one dimension representing the spatial coordinate (denoted by $x$), with a range between 0 and 100, and the other representing a (fictional) variable value (denoted by $v$), with a range between 0 and 100. The values in the two-dimensional field represent probability density. For example, the value of 0.1 at $x=23$ and $v=54$ means that the probability of the variable value at $x=23$ being between 54 and 55 is 10\%. At each $x$, the sum of the probability density over all possible $v$ values is always equal to 1. The fields are available in NetCDF format, along with the Python code that was used to generate them (see Code and data availability section for details). 

The dataset contains 20 probabilistic situations (indexed as P00, P01, P02, \ldots, P19) and 11 deterministic ones (indexed as D00, D01, D02, \ldots, D10). In deterministic situations, only a single variable value is allowed at each spatial location, with 100\% probability. 

\begin{figure}
\centering
\includegraphics[width=1\textwidth]{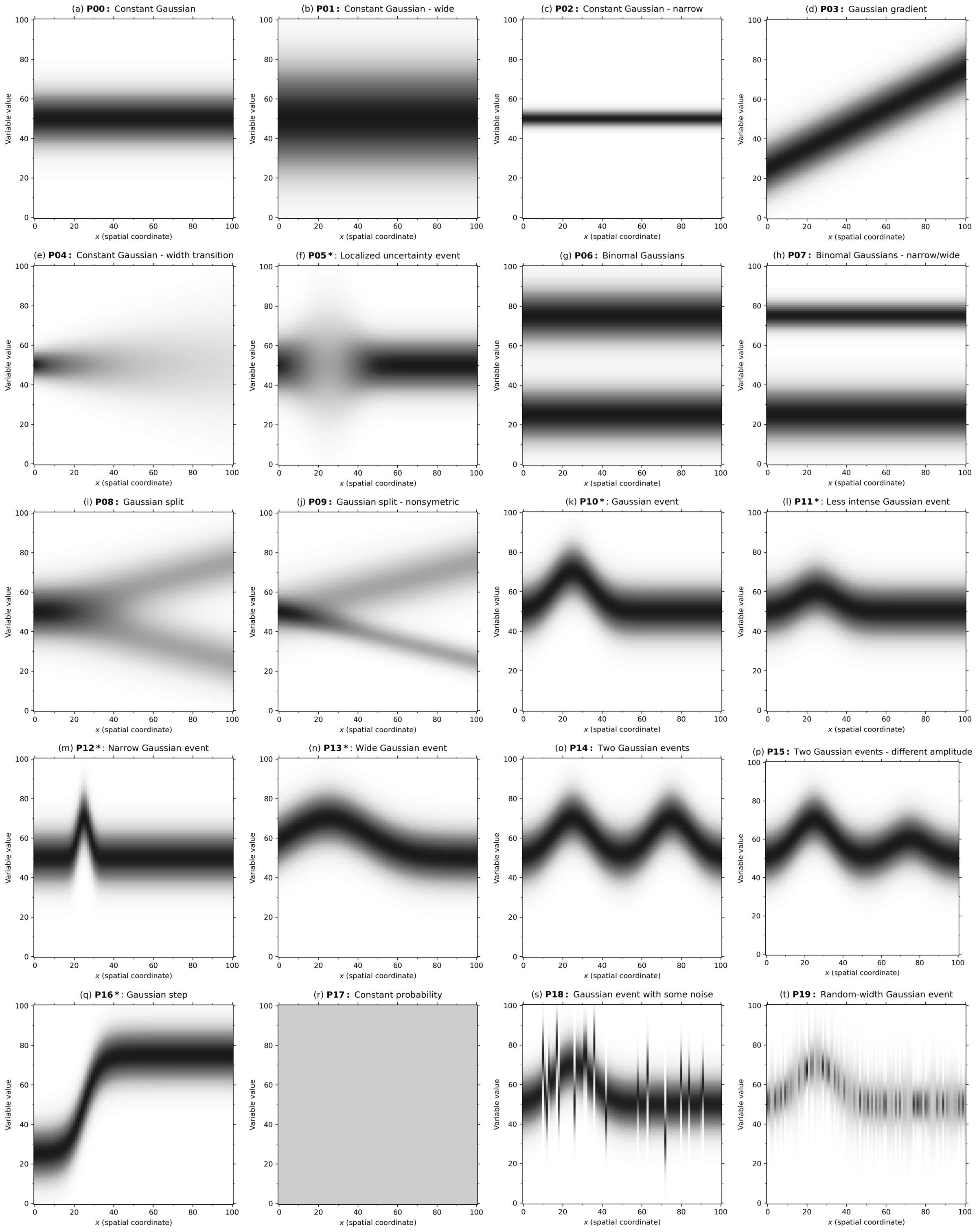}
\caption{Visualization of the idealized probabilistic situations P00--P19. The situations where the index includes a star character ("*") are available in multiple variants, with the fields spatially shifted to the right (see the text for details).}
\label{fig:fields1}
\end{figure}

\begin{figure}
\centering
\includegraphics[width=1\textwidth]{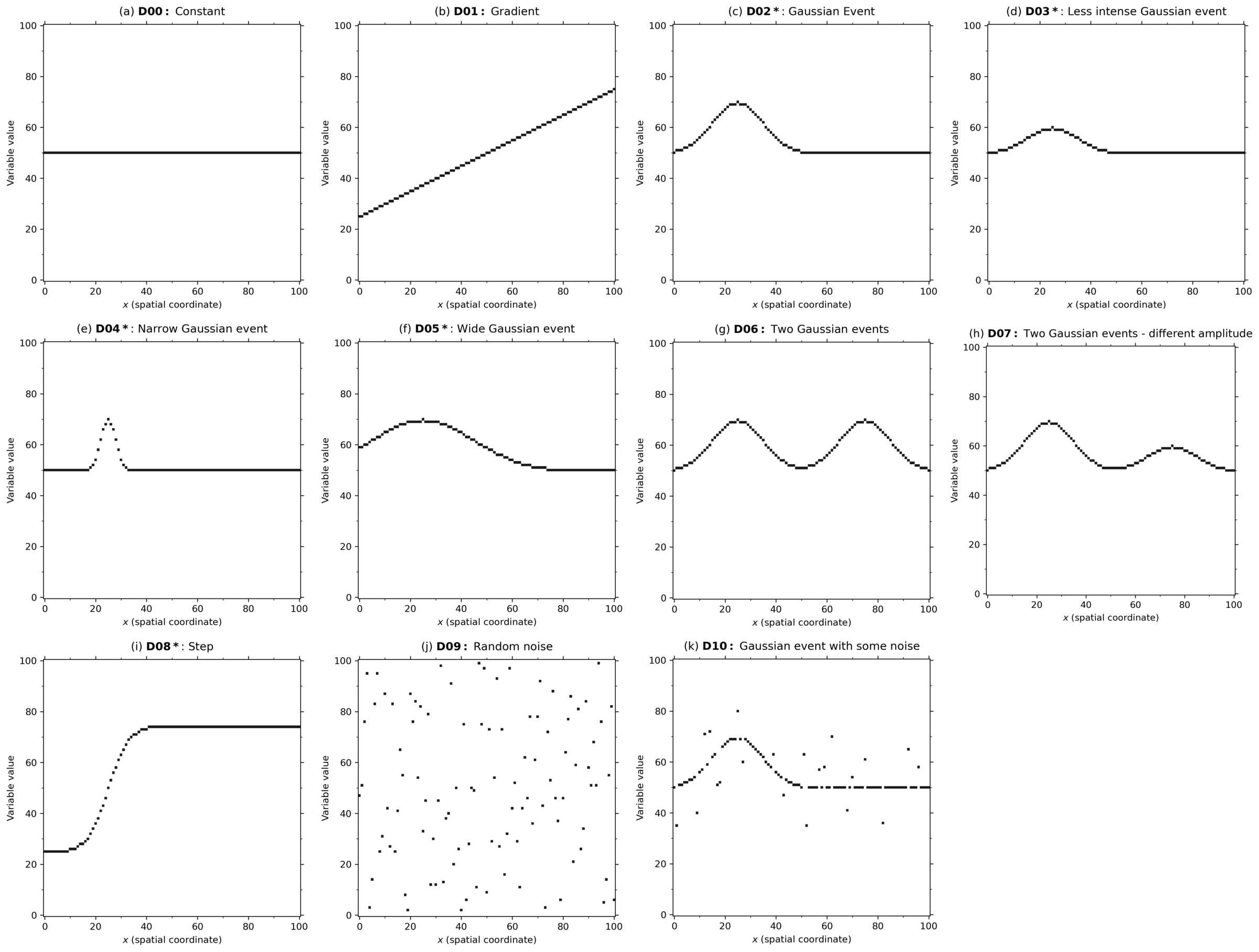}
\caption{Same as Figure~\ref{fig:fields1} but for idealized deterministic situations.}
\label{fig:fields3}
\end{figure}

The setup of the situations is quite diverse -- there are constant fields, gradients, steps, presence of noise, and events with different properties and displacements. Figures~\ref{fig:fields1}-\ref{fig:fields3} show a visualization of all the probabilistic and deterministic situations, respectively.

Some of the situations are available in multiple variants, with the fields spatially shifted toward the right (larger $x$ values). In Figures~\ref{fig:fields1}-\ref{fig:fields3}, these cases have a star character "*" shown at the end of their id. The shift introduces a spatial displacement relative to the non-shifted variant, which ranges from 0 (no shift) to 50 grid points in steps of 1 grid point. Displaced situations are denoted with a "\_XX" suffix, where XX is the shift magnitude expressed as a number of grid points. For example, P10\_05 refers to the P10 situation shifted spatially toward the right by 5 grid points. The fields can also be mirrored vertically (along the $v$) or horizontally (along the $x$), denoted with "v" or "h" suffixes, respectively. For example, P10\_05hv refers to the P10 situation shifted by 5 grid points, which has been mirrored both vertically and horizontally. 

Most of the probabilistic situations are based on a Gaussian probability distribution of variable values, which can be expressed as $A\cdot\exp{\left( -(v-v_0)^2/(2\sigma_v^2)\right)}$, where $v_0$ is the center value of the distribution and $\sigma_v$ its width (standard deviation). The amplitude $A$  is selected in such a way that the sum of the probability density over all possible $v$ values at a particular location is always equal to 1.

Situations P00--P02 (Fig.\ref{fig:fields1}(a-c)) represent spatially uniform Gaussian distributions centered at $v_0=50$ but with different widths ($\sigma_v$ is either 10, 20, or 3, respectively). Situation P03 (Fig.\ref{fig:fields1}d) represents a uniform Gaussian distribution with $\sigma_v=10$ centered at increasingly higher values towards the right, starting at $v_0=25$ on the left border (at $x=0$) and increasing linearly to $v_0=75$ on the right (at $x=100$). Situation P04 (Fig.\ref{fig:fields1}e) represents a Gaussian distribution centered at $v_0=50$ but with increasing width towards the right ($\sigma_v$ increases linearly from 5 to 20 when $x$ increases from 0 to 100). 

\begin{figure}
\centering
\includegraphics[width=1\textwidth]{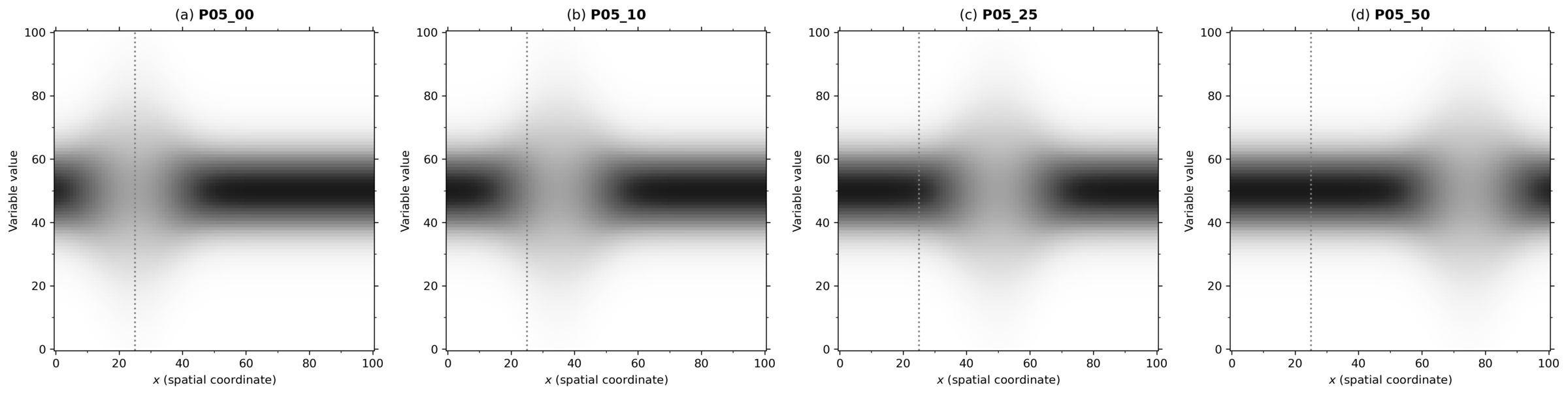}
\caption{Same as Figure~\ref{fig:fields1}f but with some spatially shifted variants of situation P05 also shown. From (a-d) the shift is either 0 (no shift), 10, 25 and 50 grid points, respectively. The dotted vertical line shows the location of the center of the original non-shifted event ($x=25$).}
\label{fig:displacement_visualization}
\end{figure}

Situation P05 (Fig.\ref{fig:fields1}f) represents a localized event of increased uncertainty. The situation is similar to P01, with a Gaussian distribution centered at $v_0=50$ and whose local change in width is modeled by using a Gaussian function as $\sigma_v(x)=10 + 10\exp{\left( -(x-x_0)^2/(2\sigma_x^2)\right)}$, where $x_0=25$ and $\sigma_x=10$. The situation is available in multiple variants, with the event shifted to the right by increasing the $x_0$ (Figure~\ref{fig:displacement_visualization}).

Situations P06--P07 (Fig.\ref{fig:fields1}(g-h)) represent bimodal distributions defined as the sum of two spatially uniform Gaussian distributions centered at $v_0=25$ and $v_0=75$. In P06, the Gaussian distributions have identical amplitude and width ($\sigma_v=10$), whereas in P07 they have identical amplitude but different widths ($\sigma_v=10$ and 5).

Situations P08--P09 (Fig.\ref{fig:fields1}(i-j)) represent a probability split, a transition from a unimodal to bimodal distribution. P08 is defined as the sum of two identical Gaussian distributions with $\sigma_v=10$ whose centers start at $v_0=50$ on the right side (at $x=0$), but with the location of the centers increasing or decreasing linearly to 75 or 25 on the right side (at $x=100$). P09 is similar to P08, but with the Gaussian distributions having identical amplitudes but different widths ($\sigma_v=10$ and 5). 

Situations P10--P13 (Fig.\ref{fig:fields1}(k-n)) represent localized events of increased intensity (increased variable value) with the centers of the Gaussian distributions moved to higher values. The local increase in $v_0$ is modeled with the help of Gaussian function as $v_0(x)=50 + b\cdot\exp{\left( -(x-x_0)^2/(2\sigma_x^2)\right)}$, where $x_0=25$, while $b$ and $\sigma_x$ depend on the situation. Similarly to P05, all situations are available in multiple variants, with the event shifted to the right. Situation P10 uses $b=20$ and $\sigma_x=10$. Compared to P10, P11 uses an event of identical width but smaller intensity ($b=10$). Contrary to P11, P12 and P13 use an event of the same intensity as P10 ($b=20$), but with either smaller or larger width ($\sigma_x=3$ and 20, respectively).

Situations P14--P15 (Fig.\ref{fig:fields1}(o-p)) include two localized events of increased intensity. The local increase in $v_0$ is modeled by the sum of two Gaussian functions as $v_0(x)=50 + b_1\cdot\exp{\left( -(x-x_1)^2/(2\sigma_x^2)\right)} + b_2\cdot\exp{\left( -(x-x_2)^2/(2\sigma_x^2)\right)}$, where $\sigma_x=10$, $x_1=25$, $x_2=75$, while $b_1$ and $b_2$ depend on the situation. In P14, the events are identical with $b_1=b_2=20$. In P15, one of the events is less intense, with $b_1=20$ and $b_2=10$.

Situation P16 (Fig.\ref{fig:fields1}q) represents a step-shaped transition event, with the variable values transitioning from lower to higher in a relatively short spatial distance. The increase in values is modeled by a logistic function as $v_0(x)=25 + 50 / \left(1 + \exp(-(x-x_0)/4)\right)$. Similarly to P05 and P10--P13, P16 is available in multiple variants, with the step event shifted to the right. 

Situation P17 (Fig.\ref{fig:fields1}r) is somewhat special in being the only probabilistic situation not based on a Gaussian distribution. The situation represents a limit case in which all values have the same probability density at all locations. 

Situations P18--P19 (Fig.\ref{fig:fields1}(s-t)) are modified versions of P10 with some randomly generated noise or variability introduced. In P18, some noise was introduced by modifying $v_0$ at 20\% of randomly chosen locations. The change in $v_0$ at each affected location was randomly sampled from the uniform distribution in the interval [-20, 20]. In P19, the $v_0$ was kept the same as in P10, but with the variability increased by randomizing the width of the Gaussian distribution. Namely, the $\sigma_v$ at each location was randomly sampled from the uniform distribution in the interval [5, 15]. 

Deterministic situations (Fig.\ref{fig:fields3}) are based on and reflect probabilistic ones. Situation D00 (Fig.\ref{fig:fields3}a) has the same value ($v=50$) everywhere. In D01 (Fig.\ref{fig:fields3}b), $v$ linearly increases from 25 on the left (at $x=0$) to 75 at the right (at $x=100$). 

Situations D02--D05 (Fig.\ref{fig:fields3}(c-f)) are analogous to P10--P13 by representing localized events of increased intensity. The local increase in $v$ is modeled by a Gaussian function as $v(x)=50 + b\cdot\exp{\left( -(x-x_0)^2/(2\sigma_x^2)\right)}$, where $x_0=25$, while $b$ and $\sigma_x$ depend on the situation. All situations are available in multiple variants, with the event shifted to the right. Situation D02 uses $b=20$ and $\sigma_x=10$. Compared to D02, D03 uses an event of identical width but smaller intensity ($b=10$). Contrary to D03, D04 and D05 use an event of the same intensity as P02 ($b=20$), but with either smaller or larger width ($\sigma_x=3$ and 20, respectively).

Situations D06--D07 (Fig.\ref{fig:fields3}(g-h)) are analogous to P14--P15 by including two localized events of increased intensity. The local increase in $v$ is modeled by the sum of two Gaussian functions and can be expressed as $v(x)=50 + b_1\cdot\exp{\left( -(x-x_1)^2/(2\sigma_x^2)\right)} + b_2\cdot\exp{\left( -(x-x_2)^2/(2\sigma_x^2)\right)}$, where $\sigma_x=10$, $x_1=25$, $x_2=75$, while $b_1$ and $b_2$ depend on the situation. In D06, the events are identical with $b_1=b_2=20$. In D07, one of the events is less intense, with $b_1=20$ and $b_2=10$. 

Situation D08 (Fig.\ref{fig:fields3}i) is analogous to P16 by representing a step-shaped transition event. The increase in value is modeled by a logistic function and can be expressed as $v(x)=25 + 50 / \left(1 + \exp(-(x-x_0)/4)\right)$. The situation is available in multiple variants, with the step event shifted to the right.

Situation D09 (Fig.\ref{fig:fields3}j) consists of random noise. Namely, at every location, the value was randomly sampled from a uniform distribution over [0, 100]. Situation D10 (Fig.\ref{fig:fields3}k) is analogous to P18 by introducing a degree of random noise to D02. More specifically, at 20\% of randomly chosen locations, the change in $v$ was randomly sampled from the uniform distribution in the interval [-20, 20].

\section{Examples of use of the dataset}\label{sec:examples}

\begin{figure}
    \centering
    \includegraphics[width=1\textwidth]{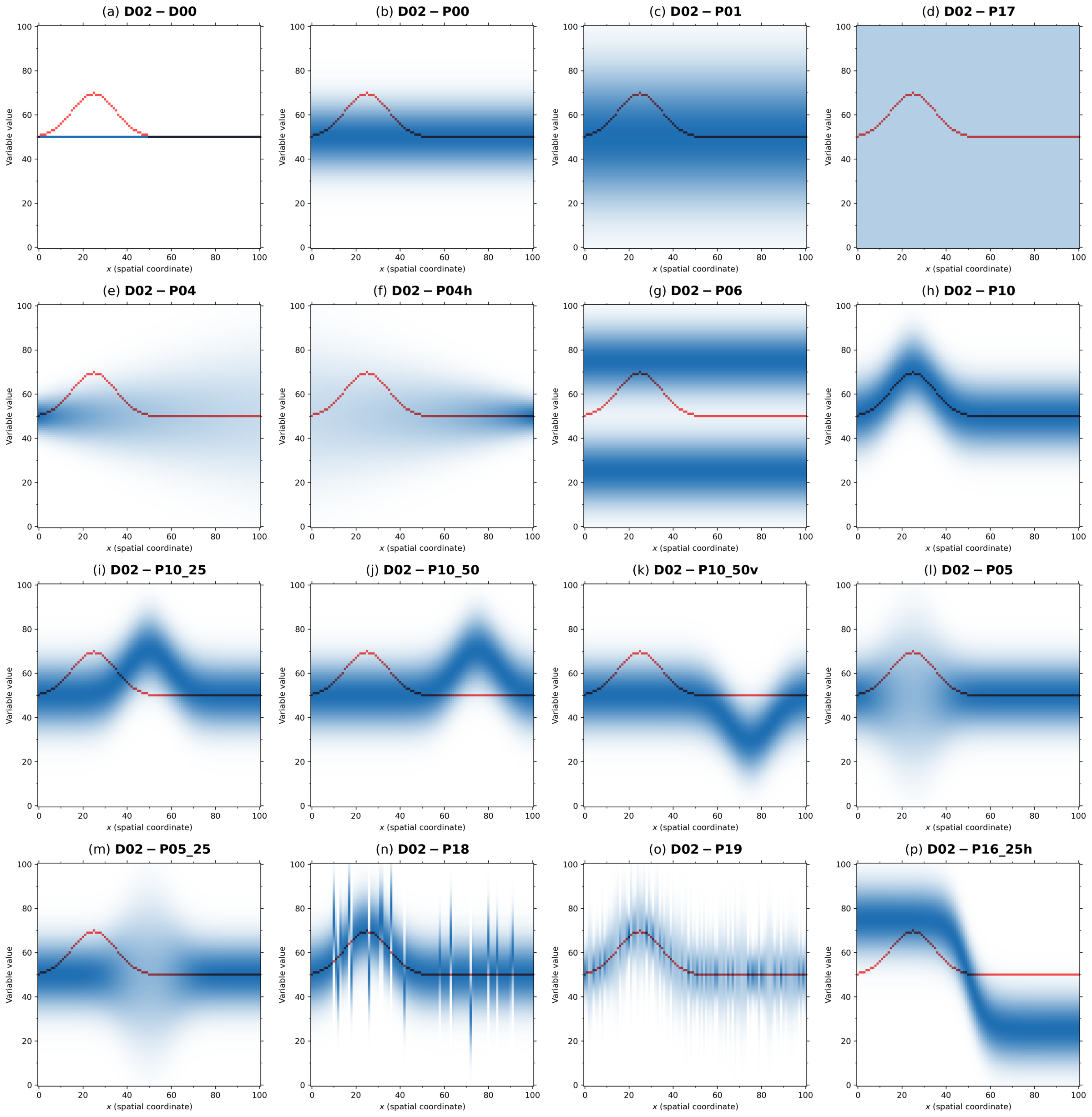}
    \caption{Visualization of the selected comparisons.}
    \label{fig:fields-comparison}
\end{figure}

\subsection{Selected comparisons}

We have selected comparisons and grouped them into five experiments to demonstrate how the dataset can be leveraged to study the behavior of verification methods. Both the relative scores associated with the different forecasts and their absolute scores can be leveraged in order to assess sensitivity to misspecifications. The variety of our dataset ensures that the same types of misspecification appear in multiple forms and that different types of misspecification are combined in order to study their joint effect.

All the forecasts are compared to the same observation: a Gaussian event (D02). Note that since the observation is a deterministic field, the score values and their expectation are the same.

\paragraph{Dispersion experiment (Figures~\ref{fig:fields-comparison}(a-g)).} The first experiment focuses on how forecast dispersion is accounted for by verification methods. We compare constant forecasts with different dispersion: probabilistic forecasts with constant Gaussian dispersion (P00-P01). Moreover, we additionally compare them to forecasts with a marginal increase in Gaussian dispersion as the spatial coordinate increases (P04) or decreases (P04h). Finally, they are also compared to a forecast with a marginal distribution with two Gaussian modes (P06).

\paragraph{Limiting experiment (Figures~\ref{fig:fields-comparison}(a,d)).} In this experiment, two limiting forecasts are compared: a deterministic constant forecast (D00) and a marginally constant forecast (P17).

\paragraph{Location experiment (Figures~\ref{fig:fields-comparison}(h-m)).} This experiment studies the impact of the misspecification of the location of the event. We compare three forecasts with a constant marginal dispersion and predicting an event of correct size and amplitude at the correct location (P10), and displaced by 25 (P10\_25) and 50 units (P10\_50), respectively. These forecasts are also compared to a forecast with a very displaced Gaussian event, but with a negative amplitude (P10\_50v). Additionally, two forecasts with a greater dispersion at the location of an event with a correct location (P05) and a displaced one (P05\_25) are included in the comparison.  

\paragraph{Noise experiment (Figures~\ref{fig:fields-comparison}(a,n-o)).} This experiment investigates the effect of noise. A Gaussian event forecast at the correct location (P10) is included as a reference. Two additional forecasts are obtained by modifying the reference forecast. A first noisy forecast has deterministic additive noise at a subset of spatial locations (P18), and the second has deterministic multiplicative noise across all spatial locations (P19), locally increasing dispersion.

\paragraph{Misfits experiment (Figures~\ref{fig:fields-comparison}(k,m,p)).} This experiment compares three very different and incorrect forecasts. It includes P10\_50v and P05\_25, which have already been used in previously introduced experiments. Moreover, it includes a forecast with a downward step as spatial coordinates increase and a constant Gaussian dispersion (P16\_25h).

\subsection{Generation of ensemble members}

With the information visualized in Figure~\ref{fig:fields1}, only the marginal probabilities of the probabilistic fields are provided. In other words, only information about the spatial variation of probability density is provided; no information is given on the joint variation of pairs of locations of the possible realizations that satisfy the provided marginal probability (i.e., spatial dependence). Such information is, for example, available in an ensemble forecast and is required to compute certain scores (e.g., the energy score). To construct a realization of an ensemble forecast from the marginal probability fields, we can consider the probabilistic fields to be second-order random fields; in this case, their spatial dependence is fully characterized by the covariance between pairs of locations. We show three cases of stationary and isotropic covariance functions (i.e., that depend only on the distance between two locations).

We first consider two exponential covariances, defined as
\begin{equation*}
    \mathrm{cov}(G(x),G(x')) = \exp\left(-\frac{\lVert x-x'\rVert}{\lambda}\right),
\end{equation*}
with a range parameter $\lambda=2000$ and $\lambda=30$, respectively, and where $x,x'\in\{0,\dots,100\}$ are locations. We later refer to such covariances as exponential covariance or simply Exp. for short. The larger the range parameter, the greater the correlation distance, which results in smoother fields.

The third covariance is an exponential covariance with a cut-off distance $c=5$ and a range parameter of $\lambda=30$. A hard cut-off would not lead to a valid covariance \citep{Guillot2012}. Hence, the covariance is obtained from an exponential covariance with values associated with pairs of locations at a distance greater than $c$ set to $0$ and then projected onto the space of positive semi-definite matrices \citep[see, e.g.,][]{Fan2016}. The cut-off ensures that there is no correlation between locations separated by a distance greater than $c$, resulting in fields that are more spatially variable. We refer to this covariance as the cut-off covariance or simply Cut-off.

\begin{figure}[h]
    \centering
    \includegraphics[width=\linewidth]{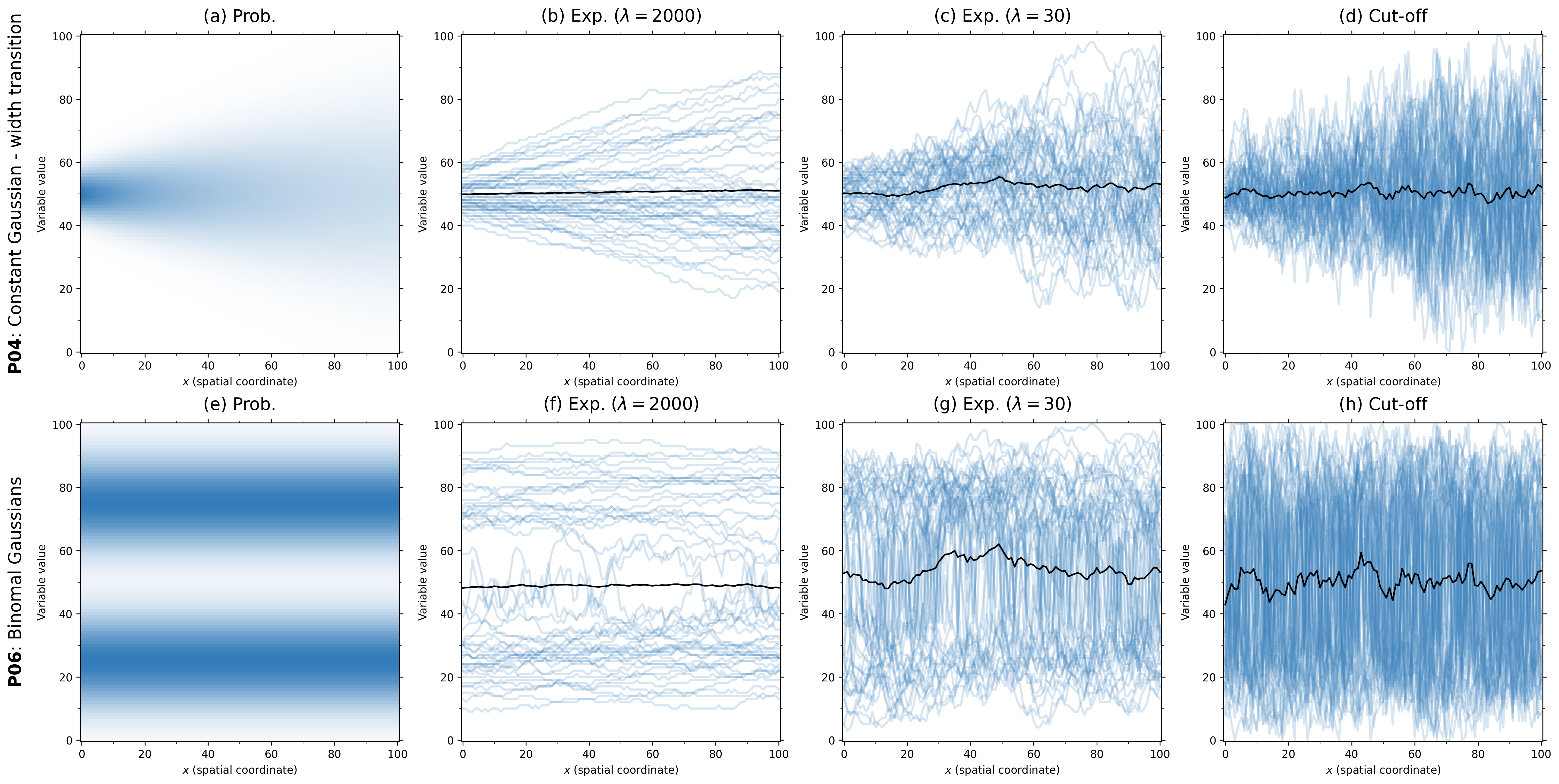}
    \caption{Illustration of the influence of the covariance. P04 (a-d) and P06 (e-h) are shown with ensembles of $M=50$ members. The two leftmost panels show the probability density fields, while the other panels show the resulting realizations of the ensemble members generated using the three variants of the covariance-based approach. The black lines correspond to the ensemble mean.}
    \label{fig:covariance}
\end{figure}

Figure~\ref{fig:covariance} illustrates the effect of the choice of covariance function. The exponential covariance with $\lambda=2000$ leads to the smoothest ensemble members (Fig.~\ref{fig:covariance}(b,f)), and the cut-off one to the roughest (Fig.~\ref{fig:covariance}(d,h)). While having the same marginal distributions, the resulting ensembles yield members with different behavior in terms of spatial smoothness and mode switching when the marginal distributions are bimodal.

\subsection{Evaluation of selected comparisons}

The aforementioned selected comparisons are evaluated using the Mean Squared Error (MSE), the aggregated Continuous Ranked Probability Score (CRPS; \cite{Matheson1976,Pic2025}), the Energy Score (ES; \cite{Gneiting2007}), and the ensemble mean Fraction Skill Score (emFSS; \cite{Roberts2008, Mittermaier2007}). Let $F$ and $y$ be a probabilistic forecast and an observation, respectively, over the spatial dimension $\{0,\dots,100\}$. The MSE is defined as
\begin{equation*}
    \mathrm{MSE}(F,y)= \frac{1}{101}\lVert \mathrm{mean}(F)-y\rVert^2 = \frac{1}{101}\sum_{x=0}^{100} (\mathrm{mean}(F_x)- y_x)^2,
\end{equation*}
where $\mathrm{mean}(\cdot)$ is the mean functional, $\lVert \cdot\rVert$ is the Euclidean norm, and $F_x$ and $y_x$ are the marginal distribution of $F$ and the value of $y$, respectively, at location $x$. The aggregated CRPS is defined as
\begin{equation*}
    \mathrm{aCRPS}(F,y)=\frac{1}{101}\sum_{x=0}^{100} \mathrm{CRPS}(F_x, y_x).
\end{equation*}
 The ES is defined as
\begin{equation*}
    \mathrm{ES}(F,y)=\mathbb{E}_F\lVert X-y\rVert - \frac{1}{2}\mathbb{E}_{F,F}\lVert X-X'\rVert,
\end{equation*}
where $X$ and $X'$ are independent random variables following $F$, and, thus, the second term measures the spread of the forecasted distribution. The emFSS is defined as
\begin{equation*}
    \mathrm{emFSS}_{b,t}(F,y)= \mathrm{FSS}_{b,t}(\mathrm{mean}(F),y),
\end{equation*}
where $\mathrm{FSS}_{b,t}$ is the standard FSS for deterministic forecasts with a neighborhood size $b$ and a threshold $t$. Here, we take $b=10$ and $t=45$ or $55$. The aggregated CRPS and MSE are grid-point-by-grid-point scores and thus non-spatial scores, as they aggregate quantities computed at each location separately. We consider only the exponential covariance with $\lambda=2000$ to generate the ensemble forecasts. As the MSE, the aggregated CRPS, and the emFSS can be computed only using the marginal distribution, we also compute their values in this setting and refer to them as Prob. in contrast to their values computed using the ensemble forecast generated by the exponential covariance, which are referred to as Exp.  We generate ensemble forecasts with 1000 members.

\FloatBarrier
\begin{table}[H]
\centering
\caption{Scores for the dispersion experiment with the observations always being D02.  For the emFSS the values are not listed in the table since its value was always $1$ for $t=45$, and 0 for $t=55$.}
\vspace{1em}
\label{tab:scores_dispersion}
\begin{tabular}{c@{\hspace{5pt}}cc|c@{\hspace{5pt}}cc|c@{\hspace{5pt}}cc|c@{\hspace{5pt}}cc|c@{\hspace{5pt}}cc}
\toprule
\multicolumn{6}{c|}{MSE} & \multicolumn{6}{c|}{Aggregated CRPS} & \multicolumn{3}{c}{ES} \\[4pt]
\# & Forecast & Prob. & \# & Forecast & Exp. & \# & Forecast & Prob. & \# & Forecast & Exp. & \# & Forecast & Exp. \\
\midrule
1 & P01 & \textbf{65.07} & 1 & P06 & \textbf{61.74} & 1 & P04h & \textbf{4.39} & 1 & P04h & \textbf{4.39} & 1 & P04h & \textbf{59.62} \\
1 & P06 & \textbf{65.07} & 2 & P01 & 63.71 & 2 & P00 & 4.48 & 2 & P00 & 4.50 & 2 & P00 & 59.78 \\
1 & P04h & \textbf{65.07} & 3 & P04h & 64.04 & 3 & P04 & 5.28 & 3 & P04 & 5.32 & 3 & P04 & 64.56 \\
1 & P04 & \textbf{65.07} & 4 & P00 & 64.50 & 4 & P01 & 5.84 & 4 & P01 & 5.90 & 4 & P01 & 70.68 \\
1 & P00 & \textbf{65.07} & 5 & P04 & 64.61 & 5 & P06 & 9.94 & 5 & P06 & 10.09 & 5 & P06 & 108.42 \\
\bottomrule
\end{tabular}
\end{table}

\paragraph{Dispersion experiment.}

Table~\ref{tab:scores_dispersion} provides the results of the dispersion experiment. In theory, all the forecasts should have the same MSE since they have the same mean (50 at each location). For Prob., this holds. However, due to the finite ensemble size, it does not hold for the dependence structure considered (Exp.). The random fluctuation of the ensemble mean for Exp. seems to benefit P06.

The aggregated CRPS accounts for forecast dispersion when comparing it to the observation. Indeed, this leads to forecasts being ranked depending on their dispersion. However, the less dispersed is not better, as forecast dispersion needs to account for the amplitude of the Gaussian event (D02).  The decreasing dispersion of P04h with increasing spatial coordinate makes it the best forecast in terms of aggregated CRPS. Furthermore, since the aggregated CRPS is a non-spatial score, the dependence structure should not affect its value. Indeed, when comparing values for Prob. and Exp., the ranking of the forecasts remains the same despite a small difference between their values.

The ES is a multivariate score that also uses the dependence structure of the forecast. It can only be computed when a dependence structure is provided. The ES ranks P04h as the best forecast for Exp., and the ranking of the competing forecasts is in perfect agreement with that of the aggregated CRPS.

Contrary to the aforementioned scores, emFSS has a positive orientation (i.e., the higher the score, the better). In theory, it does not depend on the dependence structure, as it only uses the forecast mean, and this is verified by the absence of difference between the value of Prob. and Exp. For $t=45$, all the forecast means exceed the threshold over all the neighborhoods. Thus, they all have the highest possible emFSS value of $1$. Similarly, for $t=55$, none of the forecast means exceed the threshold, and all the forecasts have the lowest possible emFSS value of $0$.

The effect of ensemble size remains present in the other experiments but will not be explicitly mentioned.

\FloatBarrier
\begin{table}[H]
\centering
\caption{Scores for the limiting experiment with the observations always being D02. For the emFSS the values are not listed in the table since its value was always $1$ for $t=45$, and 0 for $t=55$.}
\vspace{1em}
\label{tab:scores_limiting}
\begin{tabular}{c@{\hspace{5pt}}cc|c@{\hspace{5pt}}cc|c@{\hspace{5pt}}cc|c@{\hspace{5pt}}cc|c@{\hspace{5pt}}cc}
\toprule
\multicolumn{6}{c|}{MSE} & \multicolumn{6}{c|}{Aggregated CRPS} & \multicolumn{3}{c}{ES} \\[4pt]
\# & Forecast & Prob. & \# & Forecast & Exp. & \# & Forecast & Prob. & \# & Forecast & Exp. & \# & Forecast & Exp. \\
\midrule
1 & P17 & \textbf{65.07} & 1 & P17 & \textbf{61.99} & 1 & D00 & \textbf{4.63} & 1 & D00 & \textbf{4.63} & 1 & D00 & \textbf{81.07} \\
1 & D00 & \textbf{65.07} & 2 & D00 & 65.07 & 2 & P17 & 9.06 & 2 & P17 & 9.29 & 2 & P17 & 102.20 \\
\bottomrule
\end{tabular}
\end{table}

\paragraph{Limiting experiment.}
Table~\ref{tab:scores_limiting} provides the results of the limiting experiment. This experiment aims to replicate the investigation of how scores react to limiting or pathological cases. This is crucial in practice, as it can act as a sanity check or inform about the value of a score for an unskillful forecast. The MSE and the emFSS only use the forecast mean and thus rank the two forecasts equally. The emFSS has a value of 1 for $t=45$ and 0 for $t=55$. On the other hand, the aggregated CRPS and the ES agree and rank D00 as better than P17. This implies that for these scores, the greater dispersion of P17 does not capture the amplitude of the event in D02 better than the deterministic constant forecast D00.

\FloatBarrier
\begin{table}[H]
\centering
\caption{Scores for the location experiment with the observations always being D02.}
\label{tab:scores_location}
\begin{subtable}{\linewidth}
\centering
\caption{MSE, agg. CRPS and ES}
\label{tab:scores_location_main}
\begin{tabular}{c@{\hspace{5pt}}cc|c@{\hspace{5pt}}cc|c@{\hspace{5pt}}cc|c@{\hspace{5pt}}cc|c@{\hspace{5pt}}cc}
\toprule
\multicolumn{6}{c|}{MSE} & \multicolumn{6}{c|}{Aggregated CRPS} & \multicolumn{3}{c}{ES} \\[4pt]
\# & Forecast & Prob. & \# & Forecast & Exp. & \# & Forecast & Prob. & \# & Forecast & Exp. & \# & Forecast & Exp. \\
\midrule
1 & P10 & \textbf{0.19} & 1 & P10 & \textbf{0.26} & 1 & P10 & \textbf{2.34} & 1 & P10 & \textbf{2.44} & 1 & P10 & \textbf{25.19} \\
2 & P05 & 65.07 & 2 & P05 & 63.92 & 2 & P05 & 4.27 & 2 & P05 & 4.25 & 2 & P05 & 57.81 \\
2 & P05\_25 & 65.07 & 3 & P05\_25 & 64.40 & 3 & P05\_25 & 4.92 & 3 & P05\_25 & 4.95 & 3 & P05\_25 & 62.80 \\
4 & P10\_25 & 109.52 & 4 & P10\_25 & 110.04 & 4 & P10\_25 & 5.96 & 4 & P10\_25 & 5.98 & 4 & P10\_50v & 80.37 \\
5 & P10\_50 & 135.01 & 5 & P10\_50v & 132.84 & 5 & P10\_50 & 6.78 & 5 & P10\_50v & 6.69 & 5 & P10\_25 & 82.44 \\
6 & P10\_50v & 135.21 & 6 & P10\_50 & 136.23 & 6 & P10\_50v & 6.79 & 6 & P10\_50 & 6.81 & 6 & P10\_50 & 91.86 \\
\bottomrule
\end{tabular}
\end{subtable}
\begin{subtable}{\linewidth}
\centering
\caption{emFSS}
\label{tab:scores_location_emfss}
\begin{tabular}{c@{\hspace{5pt}}cc|c@{\hspace{5pt}}cc|c@{\hspace{5pt}}cc|c@{\hspace{5pt}}cc}
\toprule
\multicolumn{6}{c|}{emFSS (t=45)} & \multicolumn{6}{c}{emFSS (t=55)} \\[4pt]
\# & Forecast & Prob. & \# & Forecast & Exp. & \# & Forecast & Prob. & \# & Forecast & Exp. \\
\midrule
1 & P10 & \textbf{1.00} & 1 & P10 & \textbf{1.00} & 1 & P10 & \textbf{1.00} & 1 & P10 & \textbf{1.00} \\
1 & P10\_25 & \textbf{1.00} & 1 & P10\_25 & \textbf{1.00} & 2 & P10\_25 & 0.27 & 2 & P10\_25 & 0.27 \\
1 & P10\_50 & \textbf{1.00} & 1 & P10\_50 & \textbf{1.00} & 3 & P10\_50 & 0.00 & 3 & P10\_50 & 0.00 \\
1 & P05 & \textbf{1.00} & 1 & P05 & \textbf{1.00} & 3 & P10\_50v & 0.00 & 3 & P10\_50v & 0.00 \\
1 & P05\_25 & \textbf{1.00} & 1 & P05\_25 & \textbf{1.00} & 3 & P05 & 0.00 & 3 & P05 & 0.00 \\
6 & P10\_50v & 0.80 & 6 & P10\_50v & 0.80 & 3 & P05\_25 & 0.00 & 3 & P05\_25 & 0.00 \\
\bottomrule
\end{tabular}
\end{subtable}
\end{table}

\paragraph{Location experiment.}
Table~\ref{tab:scores_location} provides the results of the location experiment. All the scores rank P10 as the best forecast. P05 correctly predicts the location of the event but is more dispersed and includes negative variation. Thus, it is ranked as second-best by MSE, aggregated CRPS, and ES.

Forecasts with an intermediate shift (P10\_25, P05\_25) are then preferred since the location of the predicted event overlaps with the location of the observed event. P05\_25 is preferred to P10\_25 by the aggregated CRPS and the ES, as its dispersion allows for the absence of an event at this shifted location (as observed).

Forecasts with a large shift (P10\_50, P10\_50v) have similar MSE and aggregated CRPS values. For $t=45$, emFSS ranks all the forecasts equally since their mean exceeds the threshold. The only exception being P10\_50v with Exp. because the limited ensemble size seems to affect its mean. For $t=55$, P10 is the only forecast with a score of 1. P10\_25 ranks second because its mean exceeds the threshold in some neighborhoods where D02 also exceeds it, due to an overlap between the predicted and observed event locations.

\FloatBarrier
\begin{table}[H]
\centering
\caption{Scores for the noise experiment with the observations always being D02.}
\label{tab:scores_noise}
\begin{subtable}{\linewidth}
\centering
\caption{MSE, agg. CRPS and ES}
\label{tab:scores_noise_main}
\begin{tabular}{c@{\hspace{5pt}}cc|c@{\hspace{5pt}}cc|c@{\hspace{5pt}}cc|c@{\hspace{5pt}}cc|c@{\hspace{5pt}}cc}
\toprule
\multicolumn{6}{c|}{MSE} & \multicolumn{6}{c|}{Aggregated CRPS} & \multicolumn{3}{c}{ES} \\[4pt]
\# & Forecast & Prob. & \# & Forecast & Exp. & \# & Forecast & Prob. & \# & Forecast & Exp. & \# & Forecast & Exp. \\
\midrule
1 & P19 & \textbf{0.18} & 1 & P19 & \textbf{0.24} & 1 & P10 & \textbf{2.34} & 1 & P10 & \textbf{2.44} & 1 & P10 & \textbf{25.19} \\
2 & P10 & 0.19 & 2 & P10 & 0.26 & 2 & P19 & 2.36 & 2 & P19 & 2.46 & 2 & P19 & 26.25 \\
3 & P18 & 27.45 & 3 & P18 & 27.72 & 3 & P18 & 3.26 & 3 & P18 & 3.35 & 3 & P18 & 44.83 \\
\bottomrule
\end{tabular}
\end{subtable}
\begin{subtable}{\linewidth}
\centering
\caption{emFSS}
\label{tab:scores_noise_emfss}
\begin{tabular}{c@{\hspace{5pt}}cc|c@{\hspace{5pt}}cc|c@{\hspace{5pt}}cc|c@{\hspace{5pt}}cc}
\toprule
\multicolumn{6}{c|}{emFSS (t=45)} & \multicolumn{6}{c}{emFSS (t=55)} \\[4pt]
\# & Forecast & Prob. & \# & Forecast & Exp. & \# & Forecast & Prob. & \# & Forecast & Exp. \\
\midrule
1 & P10 & \textbf{1.00} & 1 & P10 & \textbf{1.00} & 1 & P10 & \textbf{1.00} & 1 & P10 & \textbf{1.00} \\
1 & P19 & \textbf{1.00} & 1 & P19 & \textbf{1.00} & 1 & P19 & \textbf{1.00} & 1 & P19 & \textbf{1.00} \\
1 & P18 & \textbf{1.00} & 1 & P18 & \textbf{1.00} & 3 & P18 & 0.98 & 3 & P18 & 0.98 \\
\bottomrule
\end{tabular}
\end{subtable}
\end{table}

\paragraph{Noise experiment.}
Table~\ref{tab:scores_noise} provides the results of the noise experiment. In theory, P10 and P19 have the same mean. However, due to the discretization of field values and boundary effects, they differ slightly. As a consequence, their MSEs are very close but not equal even for Prob. More surprisingly, the noisy forecast (P19) is preferred to the unperturbed one (P10). This difference is not perceived by the emFSSs since the forecast mean is thresholded. 

The difference in dispersion between P10 and P19 is captured by the aggregated CRPS and the ES, and it leads to P19 being ranked second, as its varying dispersion does not seem to better explain the variations in the observation. P18 exhibits additive noise that affects its mean. As a consequence, it ranks last across all scores considered.

\FloatBarrier
\begin{table}[H]
\centering
\caption{Scores for the misfits experiment with the observations always being D02.}
\label{tab:scores_misfits}
\begin{subtable}{\linewidth}
\centering
\caption{MSE, agg. CRPS and ES}
\label{tab:scores_misfits_main}
\begin{tabular}{c@{\hspace{5pt}}cc|c@{\hspace{5pt}}cc|c@{\hspace{5pt}}cc|c@{\hspace{5pt}}cc|c@{\hspace{5pt}}cc}
\toprule
\multicolumn{6}{c|}{MSE} & \multicolumn{6}{c|}{Aggregated CRPS} & \multicolumn{3}{c}{ES} \\[4pt]
\# & Forecast & Prob. & \# & Forecast & Exp. & \# & Forecast & Prob. & \# & Forecast & Exp. & \# & Forecast & Exp. \\
\midrule
1 & P05\_25 & \textbf{65.07} & 1 & P05\_25 & \textbf{64.40} & 1 & P05\_25 & \textbf{4.92} & 1 & P05\_25 & \textbf{4.95} & 1 & P05\_25 & \textbf{62.80} \\
2 & P10\_50v & 135.21 & 2 & P10\_50v & 132.84 & 2 & P10\_50v & 6.79 & 2 & P10\_50v & 6.69 & 2 & P10\_50v & 80.37 \\
3 & P16\_25h & 362.34 & 3 & P16\_25h & 358.87 & 3 & P16\_25h & 12.91 & 3 & P16\_25h & 12.82 & 3 & P16\_25h & 154.82 \\
\bottomrule
\end{tabular}
\end{subtable}
\begin{subtable}{\linewidth}
\centering
\caption{emFSS}
\label{tab:scores_misfits_emfss}
\begin{tabular}{c@{\hspace{5pt}}cc|c@{\hspace{5pt}}cc|c@{\hspace{5pt}}cc|c@{\hspace{5pt}}cc}
\toprule
\multicolumn{6}{c|}{emFSS (t=45)} & \multicolumn{6}{c}{emFSS (t=55)} \\[4pt]
\# & Forecast & Prob. & \# & Forecast & Exp. & \# & Forecast & Prob. & \# & Forecast & Exp. \\
\midrule
1 & P05\_25 & \textbf{1.00} & 1 & P05\_25 & \textbf{1.00} & 1 & P16\_25h & \textbf{0.91} & 1 & P16\_25h & \textbf{0.91} \\
2 & P10\_50v & 0.80 & 2 & P10\_50v & 0.80 & 2 & P10\_50v & 0.00 & 2 & P10\_50v & 0.00 \\
3 & P16\_25h & 0.69 & 3 & P16\_25h & 0.69 & 2 & P05\_25 & 0.00 & 2 & P05\_25 & 0.00 \\
\bottomrule
\end{tabular}
\end{subtable}
\end{table}

\paragraph{Misfits experiment.}
Table~\ref{tab:scores_misfits} provides the results of the misfits experiment. MSE, aggregated CRPS, and ES rank the forecasts in the same order: P05\_25, P10\_50v, and P16\_25h. The emFSS with $t=55$ ranks P16\_25h first. Given that it is a shifted front taking larger values at smaller spatial coordinates, there is a greater overlap over threshold exceedances compared to the other forecasts. When looking at the competing forecasts, there is no clear subjective ranking, as all forecasts mix sources of misspecification. However, such more complicated misspecification challenges our understanding of the specificity of the scores.

\subsection{Sensitivity to displacement}\label{subsec:displacement}

We illustrate the role of shifted fields in studying the sensitivity of verification methods to displacement error. This is related to investigating the double-penalty effect for ensemble forecasts. We are interested in how scores are affected by spatial shifts and how shifted forecasts compare to forecasts that do not predict the event. To that end, we use the deterministic Gaussian event D02 as an observation, and the probabilistic Gaussian event P10 and all its shifted variants as forecasts. Additionally, we use P00 as a forecast not predicting the event.

Since we are interested in how scores vary when a spatial shift is introduced, we look at relative scores defined as
\begin{equation*}
    \frac{\mathbb{E}_G[S(F_\text{s},Y)]-\mathbb{E}_G[S(F_0,Y)]}{\mathbb{E}_G[S(F_0,Y)]},
\end{equation*}
where $\mathbb{E}_G[\cdot]$ is the expectation with respect to $Y$ following $G$, and with $F_\text{s}$ and $F_0$ the shifted forecast and the non-shifted forecast, respectively. Once again, since the observations are deterministic, the expected scores equal the score values.

We compare the sensitivity of the MSE, the aggregated CRPS, and the ES to spatial shifts. We consider scores computed from the marginal distributions (Prob.) and using 1000-member ensemble forecasts with the exponential covariance with $\lambda=2000$ (Exp.) and the cut-off covariance (Cut-off). Moreover, we take the expected score of the unshifted forecast for Prob. as a reference when computing the relative scores of the MSE and the aggregated CRPS, and for Exp. for the relative scores of ES.

\begin{figure}[h]
    \centering
    \includegraphics[width=\linewidth]{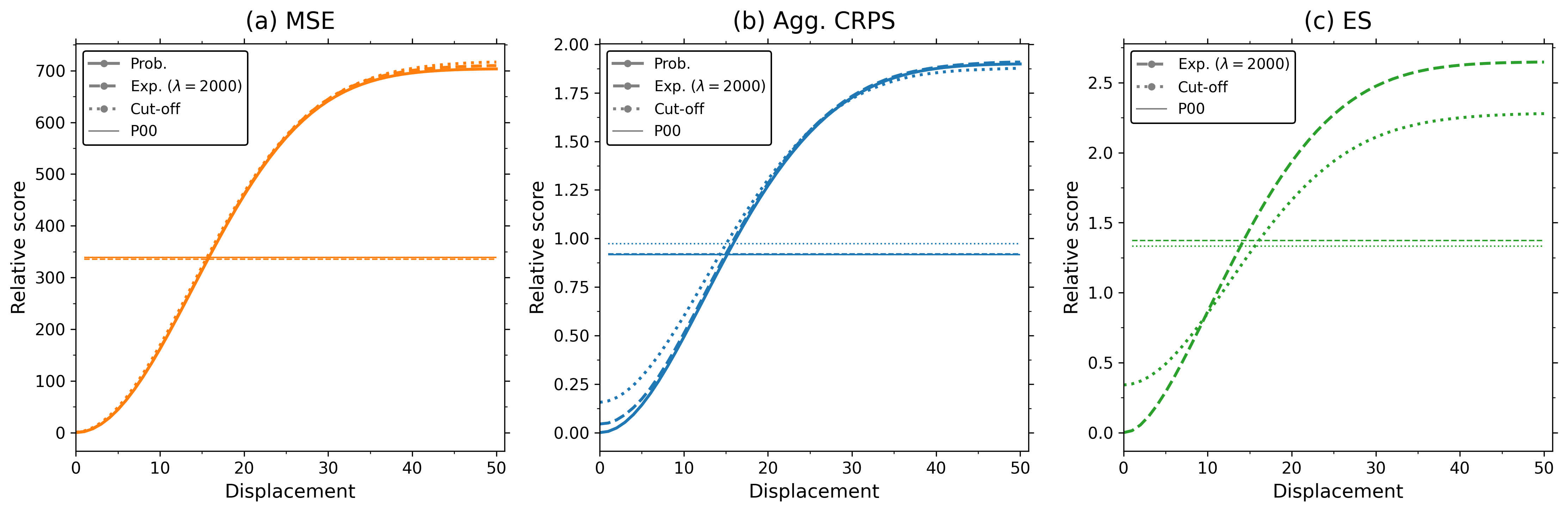}
    \caption{Relative score values for the MSE (a), the aggregated CRPS (b), and the ES (c) as spatial displacement increases for comparison between D02 and P10. The horizontal lines correspond to the comparison of D02 with P00.}
    \label{fig:sensitivity}
\end{figure}

Figure~\ref{fig:sensitivity} shows how the MSE, the aggregated CRPS, and the ES are affected by spatial shifts. All the relative scores increase with shift and appear to reach a plateau for large shifts, regardless of whether they are grid-point-by-grid-point (MSE and aggregated CRPS) or multivariate (ES).  MSE, the aggregated CRPS, and the ES seem to exhibit a similar dependence for small shifts, which is quadratic (not shown). However, the relative score of the MSE takes a greater range of values. 

The covariance seems to also affect how scores vary with shift. For MSE, the difference between Prob., Exp., and Cut-off is barely visible; this is again a consequence of the fact that MSE is only affected by the forecast mean, which itself is not affected by the spatial dependence. For the aggregated CRPS, Prob. and Exp. exhibit a very similar behavior. Cut-off, on the other hand, has a greater relative score for small shifts, but overlaps with Prob. and Exp. for larger shifts. This difference is only due to the fixed ensemble size since the aggregated CRPS is grid-point-based and insensitive to misspecifications of the spatial dependence. For ES, Exp. and Cut-off have distinct rankings. For small shifts, Exp. has a lower relative score, while for larger ones, Cut-off is better. It seems that ensemble forecasts with more spatially varying members are less affected by displacement mispecification. It may be because a greater spatial correlation (corresponding to less varying members) leads to a more correlated difference between the shifted field and the non-shifted one, and such a bigger difference between the fields may be more penalized by the ES. As Prob. can be seen as providing the theoretical value; Exp. and Cut-off provide values associated with ensemble forecasts. Hence, spatial dependence and ensemble forecasts seem to affect how scores vary with shift.

Shifted forecasts seem to have comparable relative scores as P00 for shift values of about 15 for all scores. This encourages a deeper investigation into what influences the shift value above which a forecast that does not predict the observed event is preferred.

\section{Discussion and Conclusions}

We presented an idealized probabilistic dataset composed of one-dimensional cases. It covers a wide range of probabilistic and deterministic cases: constant, localized events, gradients, fronts, noisy, and bimodal. These cases can be used to sample the spatial fields and support numerous experiments aimed at analyzing the behavior and properties of verification methods for probabilistic forecasts and comparing their behavior. Moreover, the code associated with the dataset provides great flexibility for customizing the experiments it covers. In particular, it can help investigate the double-penalty effect for probabilistic forecasts, as initiated with the experiment on the sensitivity to displacement in Section~\ref{subsec:displacement}. 

Despite the utility of the proposed dataset, it does not, by itself, cover all the cases needed to perform a complete comparison of spatial verification methods as done within ICP and MesoVICT. In particular, the two- or three-dimensional nature of weather fields increases the number of possible misspecifications. For example, anisotropy can only be investigated in at least two dimensions. Nevertheless, this dataset represents the first building block of the more extensive comparison dataset of the \textit{Bridging The Gap} project, which aims to facilitate the development and comparison of spatial verification methods for probabilistic forecasts \citep{BTG2026}.

\section*{Code and data availability}
The dataset and code associated with this article are openly available in the Zenodo repository at \url{https://doi.org/10.5281/zenodo.22836422} \citep{zenodo_repository} as well as github at \url{https://github.com/pic-romain/1d-idealized-probabilistic-fields}.

\section*{Author Contributions}
Both authors contributed equally to this work.

\section*{Acknowledgements}
The authors are grateful to Llorenç Lledó (European Centre for Medium-Range Weather Forecasts, Bonn, Germany) for a fruitful discussion regarding the development of the new dataset. 

\section*{Conflict of Interest Statement}
The authors declare no conflicts of interest.

\section*{Funding}
This research was supported by the Slovenian Research And Innovation Agency (Javna agencija za znanstvenoraziskovalno in inovacijsko dejavnost RS) research core funding No. P1-0188. This research was further supported by the University of Ljubljana Grant SN-ZRD/22-27/0510. Romain Pic gratefully acknowledges funding from the NCCR CLIM+.

\bibliography{references}  

@article{Skok2022,
author = {Skok, Gregor},
doi = {10.3390/app12084048},
issn = {2076-3417},
journal = {Applied Sciences},
month = {apr},
number = {8},
pages = {4048},
title = {{A New Spatial Distance Metric for Verification of Precipitation}},
url = {https://www.mdpi.com/2076-3417/12/8/4048},
volume = {12},
year = {2022}
}

@article{Gilleland2020,
author = {Gilleland, Eric and Skok, Gregor and Brown, Barbara G. and Casati, Barbara and Dorninger, Manfred and Mittermaier, Marion P. and Roberts, Nigel and Wilson, Laurence J.},
doi = {10.1175/MWR-D-19-0256.1},
issn = {0027-0644},
journal = {Monthly Weather Review},
month = {apr},
number = {4},
pages = {1653--1673},
title = {{A Novel Set of Geometric Verification Test Fields with Application to Distance Measures}},
url = {http://journals.ametsoc.org/doi/10.1175/MWR-D-19-0256.1},
volume = {148},
year = {2020}
}

@article{Roberts2008,
author = {Roberts, Nigel M. and Lean, Humphrey W.},
doi = {10.1175/2007MWR2123.1},
issn = {1520-0493},
journal = {Monthly Weather Review},
month = {jan},
number = {1},
pages = {78--97},
title = {{Scale-Selective Verification of Rainfall Accumulations from High-Resolution Forecasts of Convective Events}},
url = {http://journals.ametsoc.org/doi/10.1175/2007MWR2123.1},
volume = {136},
year = {2008}
}

@article{Gilleland2017,
author = {Gilleland, Eric},
doi = {10.1175/WAF-D-16-0134.1},
issn = {15200434},
journal = {Weather and Forecasting},
number = {1},
pages = {187--198},
title = {{A new characterization within the spatial verification framework for false alarms, misses, and overall patterns}},
volume = {32},
year = {2017}
}

@incollection{Brown2011,
address = {Chichester, UK},
author = {Brown, Barbara G. and Gilleland, Eric and Ebert, Elizabeth E.},
booktitle = {Forecast Verification},
doi = {10.1002/9781119960003.ch6},
month = {feb},
pages = {95--117},
publisher = {John Wiley and Sons, Ltd},
title = {{Forecasts of Spatial Fields}},
url = {https://onlinelibrary.wiley.com/doi/10.1002/9781119960003.ch6},
year = {2012}
}

@article{Dorninger2018,
author = {Dorninger, Manfred and Gilleland, Eric and Casati, Barbara and Mittermaier, Marion P. and Ebert, Elizabeth E. and Brown, Barbara G. and Wilson, Laurence J.},
doi = {10.1175/BAMS-D-17-0164.1},
issn = {00030007},
journal = {Bulletin of the American Meteorological Society},
number = {9},
pages = {1887--1906},
title = {{The setup of the MesoVICT project}},
volume = {99},
year = {2018}
}

@article{Gilleland2009,
   author = {Eric Gilleland and David Ahijevych and Barbara G. Brown and Barbara Casati and Elizabeth E. Ebert},
   doi = {10.1175/2009WAF2222269.1},
   issn = {1520-0434},
   issue = {5},
   journal = {Weather and Forecasting},
   month = {10},
   pages = {1416-1430},
   title = {Intercomparison of Spatial Forecast Verification Methods},
   volume = {24},
   year = {2009},
}

@article{Skok2025,
   author = {Gregor Skok and Llorenç Lledó},
   doi = {10.1002/qj.5006},
   issn = {0035-9009},
   journal = {Quarterly Journal of the Royal Meteorological Society},
   month = {5},
   title = {Spatial verification of global precipitation forecasts},
   year = {2025}
}

@article{Pic2025,
  title = {Proper scoring rules for multivariate probabilistic forecasts based on aggregation and transformation},
  volume = {11},
  issn = {2364-3587},
  url = {http://dx.doi.org/10.5194/ascmo-11-23-2025},
  doi = {10.5194/ascmo-11-23-2025},
  number = {1},
  journal = {Advances in Statistical Climatology, Meteorology and Oceanography},
  publisher = {Copernicus GmbH},
  author = {Pic, Romain and Dombry, Clément and Naveau, Philippe and Taillardat, Maxime},
  year = {2025},
  month = Mar,
  pages = {23–58},
}

@article{Matheson1976,
  author = {James E. Matheson and Robert L. Winkler},
  journal = {Management Science},
  title = {Scoring Rules for Continuous Probability Distributions},
  year = {1976},
  volume = {22},
  doi = {10.2307/2629907},
  issue = {10},
  page = {1087--1096},
}

@article{Mittermaier2007,
  title = {Improving short‐range high‐resolution model precipitation forecast skill using time‐lagged ensembles},
  volume = {133},
  issn = {1477-870X},
  url = {http://dx.doi.org/10.1002/qj.135},
  doi = {10.1002/qj.135},
  number = {627},
  journal = {Quarterly Journal of the Royal Meteorological Society},
  publisher = {Wiley},
  author = {Mittermaier, Marion P.},
  year = {2007},
  month = July,
  pages = {1487–1500},
}

@Article{Gneiting2007,
  author    = {Gneiting, Tilmann and Raftery, Adrian E},
  journal   = {Journal of the American Statistical Association},
  title     = {Strictly Proper Scoring Rules, Prediction, and Estimation},
  year      = {2007},
  issn      = {1537-274X},
  month     = mar,
  number    = {477},
  pages     = {359--378},
  volume    = {102},
  doi       = {10.1198/016214506000001437},
  publisher = {Informa UK Limited},
}

@article {Ahijevych2009,
      author = "David  Ahijevych and Eric  Gilleland and Barbara G.  Brown and Elizabeth E.  Ebert",
      title = "Application of Spatial Verification Methods to Idealized and NWP-Gridded Precipitation Forecasts",
      journal = "Weather and Forecasting",
      year = "2009",
      publisher = "American Meteorological Society",
      address = "Boston MA, USA",
      volume = "24",
      number = "6",
      doi = "10.1175/2009WAF2222298.1",
      pages=      "1485 - 1497",
      url = "https://journals.ametsoc.org/view/journals/wefo/24/6/2009waf2222298_1.xml"
}

@article{Ebert2008,
author = {Ebert, Elizabeth E.},
title = {Fuzzy verification of high-resolution gridded forecasts: a review and proposed framework},
journal = {Meteorological Applications},
volume = {15},
number = {1},
pages = {51-64},
doi = {https://doi.org/10.1002/met.25},
url = {https://rmets.onlinelibrary.wiley.com/doi/abs/10.1002/met.25},
eprint = {https://rmets.onlinelibrary.wiley.com/doi/pdf/10.1002/met.25},
year = {2008}
}

@article{Casati2022,
      author = "Barbara Casati and Manfred Dorninger and Caio A. S. Coelho and Elizabeth E. Ebert and Chiara Marsigli and Marion P. Mittermaier and Eric Gilleland",
      title = "The 2020 International Verification Methods Workshop Online: Major Outcomes and Way Forward",
      journal = "Bulletin of the American Meteorological Society",
      year = "2022",
      publisher = "American Meteorological Society",
      address = "Boston MA, USA",
      volume = "103",
      number = "3",
      doi = "10.1175/BAMS-D-21-0126.1",
      pages=      "E899 - E910",
      url = "https://journals.ametsoc.org/view/journals/bams/103/3/BAMS-D-21-0126.1.xml"
}

@Article{Lang2026,
    author={Lang, Simon and Alexe, Mihai and Clare, Mariana C. A. and Roberts, Christopher and Adewoyin, Rilwan and Ben Bouall{\`e}gue, Zied and Chantry, Matthew and Dramsch, Jesper and Dueben, Peter D. and Hahner, Sara and Maciel, Pedro and Prieto-Nemesio, Ana and O'Brien, Cathal and Pinault, Florian and Polster, Jan and Raoult, Baudouin and Tietsche, Steffen and Leutbecher, Martin},
    title={AIFS-CRPS: ensemble forecasting using a model trained with a loss function based on the continuous ranked probability score},
    journal={npj Artificial Intelligence},
    year={2026},
    month={Feb},
    day={02},
    volume={2},
    number={1},
    pages={18},
    issn={3005-1460},
    doi={10.1038/s44387-026-00073-7},
    url={https://doi.org/10.1038/s44387-026-00073-7}
}

@Article{Price2025,
    author={Price, Ilan
    and Sanchez-Gonzalez, Alvaro
    and Alet, Ferran
    and Andersson, Tom R.
    and El-Kadi, Andrew
    and Masters, Dominic
    and Ewalds, Timo
    and Stott, Jacklynn
    and Mohamed, Shakir
    and Battaglia, Peter
    and Lam, Remi
    and Willson, Matthew},
    title={Probabilistic weather forecasting with machine learning},
    journal={Nature},
    year={2025},
    month={Jan},
    day={01},
    volume={637},
    number={8044},
    pages={84-90},
    issn={1476-4687},
    doi={10.1038/s41586-024-08252-9},
    url={https://doi.org/10.1038/s41586-024-08252-9}
}

@article{Fan2016,
  title = {An overview of the estimation of large covariance and precision matrices},
  volume = {19},
  issn = {1368-423X},
  url = {http://dx.doi.org/10.1111/ectj.12061},
  doi = {10.1111/ectj.12061},
  number = {1},
  journal = {The Econometrics Journal},
  publisher = {Oxford University Press (OUP)},
  author = {Fan, Jianqing and Liao, Yuan and Liu, Han},
  year = {2016},
  month = Feb,
  pages = {C1–C32},
}

@article{Guillot2012,
title = {Retaining positive definiteness in thresholded matrices},
journal = {Linear Algebra and its Applications},
volume = {436},
number = {11},
pages = {4143-4160},
year = {2012},
issn = {0024-3795},
doi = {https://doi.org/10.1016/j.laa.2012.01.013},
url = {https://www.sciencedirect.com/science/article/pii/S0024379512000614},
author = {Dominique Guillot and Bala Rajaratnam},
}

@misc{BTG2026,
  author = {{Bridging The Gap}},
  title = {Bridging The Gap website},
  year = {2026},
  note = {Accessed: 2026-09-18},
  howpublished = {\url{https://pic-romain.github.io/bridging-the-gap/}},
}

@software{zenodo_repository,
  author       = {Romain Pic and Gregor Skok},
  title        = {1D Idealized Probabilistic Fields},
   year         = 2026,
  publisher    = {Zenodo},
  doi          = {10.5281/zenodo.22836422},
  url          = {https://doi.org/10.5281/zenodo.22836422},
}

\end{document}